\documentclass{aastex}
\usepackage{emulateapj5,apjfonts}

\def\lsim{\mathrel{\rlap{\lower4pt\hbox{\hskip1pt$\sim$}}
    \raise1pt\hbox{$<$}}}                % less than or approx. symbol
    
\shorttitle{X-ray point sources in galaxies}
\shortauthors{Kilgard et al.}

\begin{document}

\title{A Mini-Survey of X-Ray Point Sources in Starburst and
Non-Starburst Galaxies}

\author{R.\ E.\ Kilgard, P.\ Kaaret, M.\ I.\ Krauss, A.\ H.\
Prestwich, M.\ T.\ Raley, A.\ Zezas}

\affil{Harvard-Smithsonian Center for Astrophysics,
Cambridge, MA 02138}

\begin{abstract}

We present a comparison of X-ray point source luminosity
functions of 3 starburst galaxies (the Antennae, M82, and NGC
253) and 4 non-starburst spiral galaxies (NGC 3184, NGC 1291, 
M83, and IC 5332).  We find that the luminosity functions of
the starbursts are flatter than those of the spiral galaxies;
the starbursts have relatively more sources at high
luminosities.  This trend extends to early-type galaxies
which have steeper luminosity functions than spirals.  We
show that the luminosity function slope is correlated with
$60 \mu \rm m$ luminosity, a measure of star formation.  We
suggest that the difference in luminosity functions is
related to the age of the X-ray binary populations and
present a simple model which highlights how the shape of the
luminosity distribution is affected by the age of the
underlying X-ray binary population.

\end{abstract}

\keywords{surveys --- galaxies: spiral ---
galaxies: starburst --- X-rays: galaxies --- X-rays: sources}

\section{Introduction}

The study of X-ray point source populations in external
galaxies may permit insights into the evolution of various
stellar populations within galaxies \citep{fab89}.  Since most
luminous X-ray point sources are related to objects produced
at the end-point of the evolution of massive stars, the
population of bright X-ray point sources should be related to
the history of massive star formation.  With the recent
launch of the Chandra and XMM-Newton X-ray observatories, we
can finally study the X-ray point source populations in a
sample of galaxies with a variety of star formation histories
with high angular resolution and good sensitivity.  This may
permit insights into the relation of the X-ray point source
populations to the stellar evolutionary history of each
galaxy.

There is evidence to suggest that galaxies with high star
formation rates contain a greater number of ultra-luminous
X-ray sources ($L_x >10^{39} \rm \, erg \, s^{-1}$).  For
example, there are several such sources in each of the
starburst galaxies M82 \citep{zezas01}, the Antennae
\citep{fab01}, NGC 3256 \citep{lira01}.  By contrast, none
are found in the non-starburst/normal spiral galaxies IC 5332
or NGC 3184 (see Fig.~\ref{fig:lfs}).  However, several
biases must be considered that could affect such
``anecdotal'' evidence.  More massive galaxies may have more
ultra-luminous X-ray sources because there is a greater
likelihood of finding rare sources in larger systems.  Nearby
galaxies (e.g.\ Local Group galaxies) have such large angular
extents that Chandra surveys have been incomplete, and
sources in unexplored regions may have been missed.  The most
important biases are probably source confusion and higher
source detection thresholds in more distant galaxies.

Here we study the X-ray point source luminosity functions
(LFs) in a ``mini-survey'' of Chandra observations of four
spiral and three starburst galaxies.  Our sample covers a
range of Hubble types and concentrates on nearby galaxies to
reduce problems with confusion and incompleteness.  We
describe our sample of galaxies and data analysis procedures
in \S 2.  Our results on the X-ray luminosity
functions are presented in \S 3 and compared with previous
results in \S 4.  In \S 5, we describe a simple model which
attempts to relate the luminosity function to the age of the
X-ray binary population in each galaxy.  We present our
conclusions in \S 6.

\section{Sample Selection and Data Analysis}

We chose a sample of nearby galaxies representing many
different Hubble types in an attempt to reduce the intrinsic
biases and study the dependence of the LF parameters on star
formation activity.  For the spirals, we selected relatively
nearby galaxies with favorable inclinations in order to
minimize problems with source confusion, incompleteness, and
obscuration.  We obtained new Chandra observations for NGC
1291 and IC 5332 and used archival data for NGC 3184 and M83.
An analysis of NGC 1291 has also been presented by
\citet{irwin01}. The paucity of nearby starbursts forces us
to relax our standards.  We selected the two nearby
starbursts with available Chandra data M82 (Griffiths et al.\
2001; Zezas et al.\ 2001; Ward et al., in preparation)
and NGC 253 \citep{strickland00}.  To increase the sample, we
also include the more distant Antennae galaxies
\citep{fab01,zezas01a}.  More than 70\% of the $D_{25}$
ellipse of each galaxy, except one, was in the Chandra field
of view.  The exception was NGC 253 where the field of view
covered only 45\% of the $D_{25}$ ellipse, but did include
the entire starburst region.

\begin{figure} \epsscale{0.5} \plotone{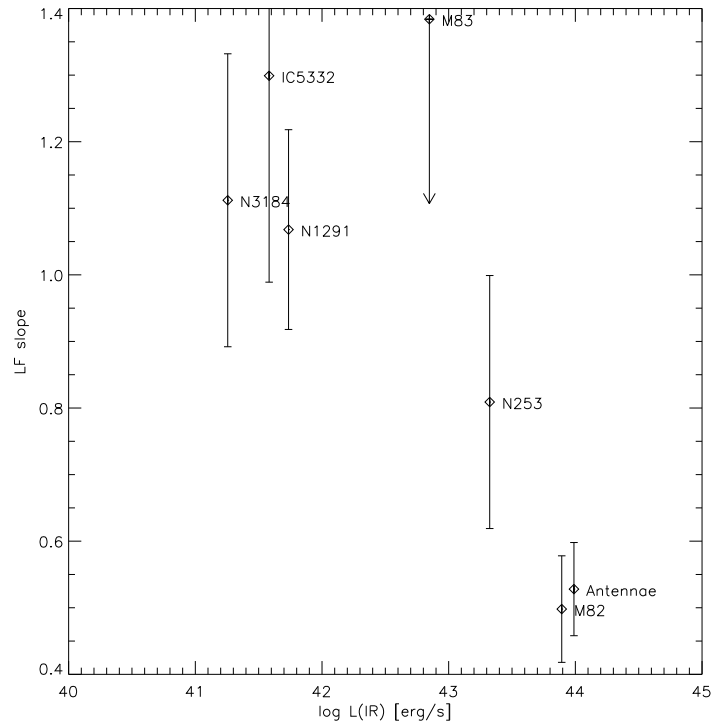}
\caption{Cumulative luminosity function slope vs. 60 \micron\ luminosity. \label{fig:LF-IR}}   \end{figure}

We note that M83 has a starburst in the nuclear and bar
region (e.g. \citet{telesco93}), but the star formation rate
is low compared to classic starbursts \citep{rieke78} and
confined to a small part of the galaxy.  Most of our detected
X-ray sources are outside the starburst region (approximately
30 arcsecs around the nucleus; \citet{alton98}) and are
representative of the disk population rather than the
starburst population.   M83 is probably best characterized,
in general, as a composite object.  However, our luminosity
function for this galaxy is dominated by the disk population
and we group it together with the spirals.  The 60 \micron\
luminosity plotted in figure~\ref{fig:LF-IR} is an upper
limit to the luminosity from the area over which we detect
the most X-ray sources.

Observations of all galaxies (except M82) were performed on
the back-illuminated ACIS-S3 CCD.  M82 was observed on the
ACIS-I3 front-illuminated CCD, which has lower sensitivity
below 2 keV than ACIS-S3.  However, the difference in sensitivity is not
significant compared with the uncertainties in distances and
assumed source spectra.

All analyses were performed using the Chandra Interactive
Analysis of Observations software package ({\it CIAO}) v2.1.2
and the Chandra Calibration Database v2.6.  Data were
screened for times of high background, as many of the
observations occurred near periods of solar activity.
Exposure times of the screened observations are in
Table~\ref{tab:prop}.  Source lists were constructed using
{\it wavdetect} \citep{pef02}, the Mexican-hat wavelet source
detection routine which is part of {\it CIAO}.  We ran {\it
wavdetect} on 0.3-6.5~keV band images using wavelet scales of
2, 4, 8, and 16 pixels.  This combination of energy band and
wavelet scales yielded the fewest spurious detections and
found all visually obvious point sources.  The output source
regions were visually inspected to remove the  artifacts that
sometimes occur with wavdetect: regions where the ellipse
minor axis is  zero, regions containing 2 sources, and
sources detected twice.  Nuclear sources, when present, were
removed from the source lists, as were sources outside the
$D_{25}$ ellipses.

Source regions were taken to be 4 times the standard deviation of the
distribution.  This over-estimates the source size, but the
contribution from background is negligible.
For each source, PHA spectra were extracted and RMFs and ARFs
were constructed.  Corresponding background spectra were
extracted for each source.  The region for each background
spectrum was taken to be an ellipse with radii equal to 4
times the source radii and excluding the source regions and
any other overlapping source regions.  To limit contamination
from any diffuse emission, the background ellipse radii were
not allowed to exceed 50 pixels (about 25\arcsec).  We
estimated fluxes for the 0.3-8~keV band from the spectrum of
each source using {\it Sherpa} and assuming two standard
models: a 5~keV thermal bremsstrahlung model with
photoelectric absorption and a power-law with photon index of
1.5 and photoelectric absorption.  In both cases, the n(H)
was fixed at the Galactic value.  Since most of the sources
do not have enough counts for spectral analysis, all model
parameters were frozen except for the normalizations.  The
0.3-8~keV fluxes were used to calculate luminosities, using
distances from the Nearby Galaxies Catalog \citep{tully88}
which assumes $H_o=75$ km s$^{-1}$ Mpc$^{-1}$ coupled
with a Virgo infall model.  Below, we
present results using the thermal bremsstrahlung model;
however, the results are essentially unchanged if the
power-law model is used.  A more detailed description of the
data analysis, discussing the relevant biases and presenting
source lists with spectral and timing results will be
presented in a future publication (Kilgard et al., in
preparation).

\section{Luminosity Functions of Spiral and Star Forming Galaxies}
\label{sec:lfs}

\begin{figure} \plottwo{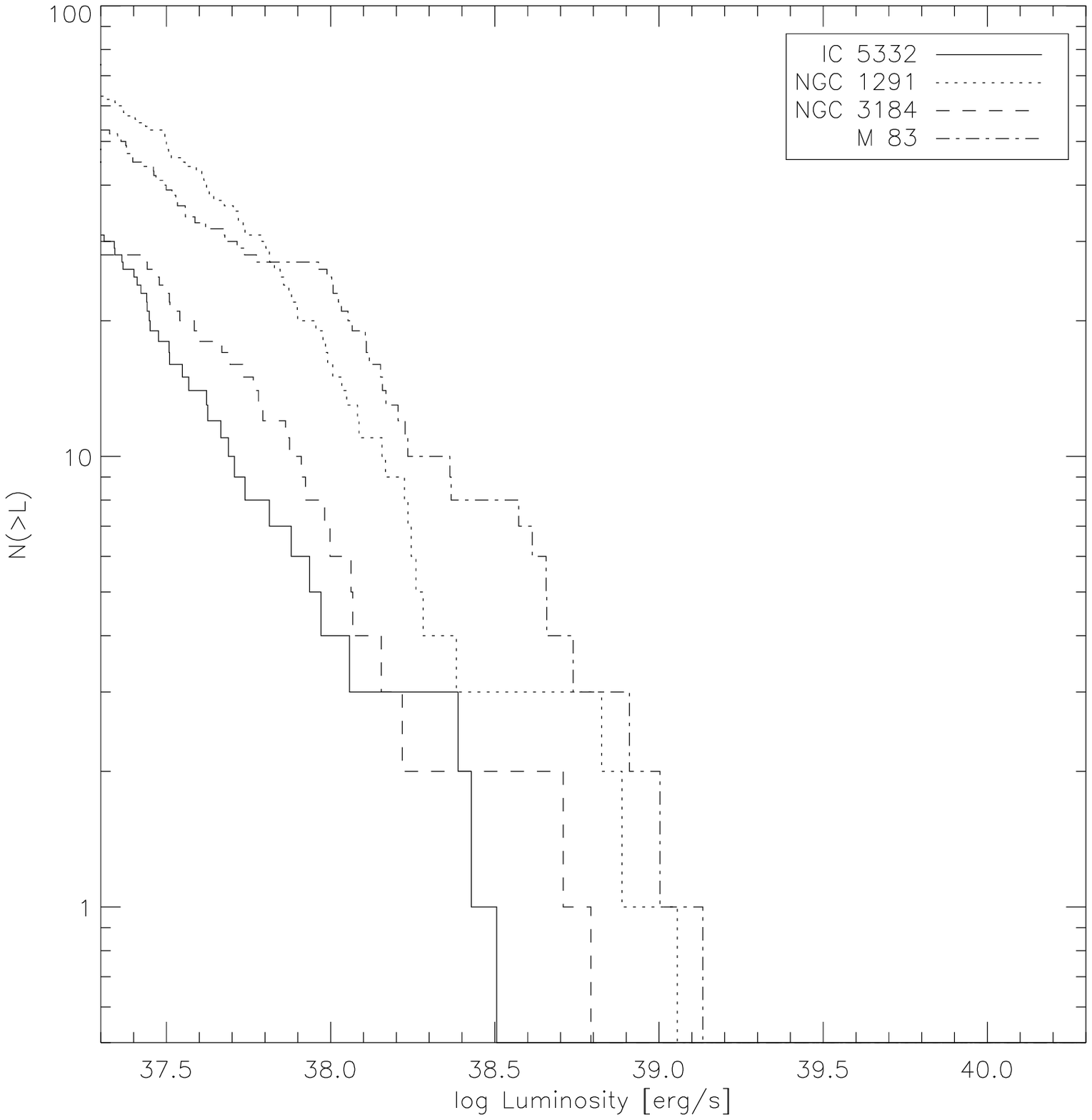}{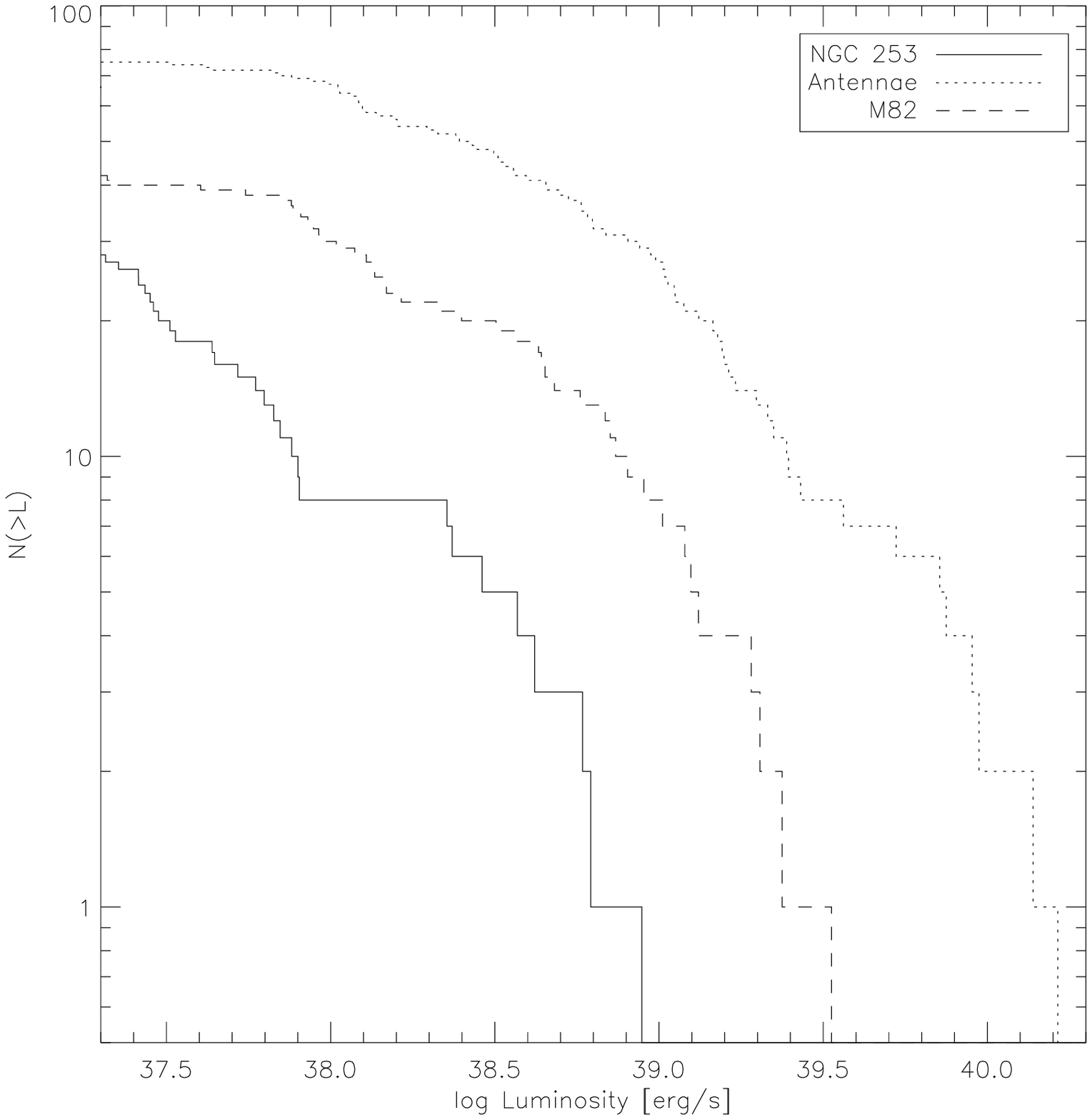} \caption{Cumulative
Luminosity functions of spirals (left) and starbursts
(right).\label{fig:lfs}} \end{figure}

The cumulative luminosity functions for all galaxies are
shown in Figure~\ref{fig:lfs}.  In order to avoid incompleteness, we
consider only the high
luminosity range of the luminosity functions.   For the nearest
galaxies in our sample, we set an overall conservative completeness
limit at the luminosity corresponding to a detection limit of 10
counts for NGC 1291, $L > 3 \times 10^{37} \rm \, erg \, s^{-1}$.  
In the case of the Antennae, this detection limit translates to $L > 9
\times 10^{37} \rm \, erg \, s^{-1}$.
We derived power-law fits to the high luminosity
range of the unbinned differential luminosity function
using a maximum likelihood statistic following \citet{crawford70}.
The slopes range between 0.50 and 1.30; the fitted slopes, $\gamma$, are
shown in Table~\ref{tab:prop}.  We evaluated the goodness of fit using
a binned differential 
luminosity distribution and calculating the $\chi^{2}$\ statistic with
errors computed following Gehrels (1986).  The reduced $\chi^{2}$\ is
between 0.2 and 1.0 for all galaxies except M83, which has a value of
1.6.  M83 shows evidence for a break in the LF at $\sim
9\times10^{37}$ ergs s$^{-1}$.  We therefore refit the power-law
including only data above this break and use this value in
Table~\ref{tab:prop}.  The reduced $\chi^{2}$\ for the new fit is 0.96,
indicating that a single power-law is an adequate model in this
luminosity range.

Since the objects in our sample extend over large areas, the
effect of interlopers in the LFs may be important. In order
to assess this effect, we estimated the distribution of
background sources using the logN-logS relation of
\citet{giacconi01}. Scaling the function according to the 
exposure time for each of our observations, we modeled the luminosity
distribution of the sources with a power-law.  We used this component
as background for our LFs and we performed the fits again.
There was little difference (less than $1\sigma$) in the results of
these fits and those presented above.

%The LFs of M82, M83, and the Antennae in fig~\ref{fig:lfs} show
%evidence for a break at $L = 6 \times 10^{37} \rm \, erg \, s^{-1}, 9
%\times 10^{37} \rm \, erg \, s^{-1}, and 10^{39} \rm \, erg \,
%s^{-1}$, respectively.  However, given the excellent fits with the
%single power-law and the large uncertainties in our data points, we do
%not consider an more complicated models (such as broken power-laws).
%Moreover, these three galaxies exhibit high levels of diffuse emission
%which varies over the surveyed area.  This causes variations of the
%detection limits over the galaxy which may result in apparent breaks
%in the discrete source LFs \citep{zezas01a}.

The early-type spiral NGC 1291 has been studied in detail by
\citet{irwin01}, who considered only bulge sources and found
no sources above $L = 2.5 \times 10^{38} \rm \, ergs \,
s^{-1}$.  We include all sources in the galaxy within the S3
chip (which covers $>$80\% of the total area of the galaxy),
which adds some higher luminosity sources.  It is possible
that in NGC 1291 we are seeing two populations (bulge and
disk) with the disk sources becoming more prominent at higher
luminosities.  However, there are only 2 disk sources
brighter than  $ L > 3 \times 10^{38} \rm \, ergs \, s^{-1}$,
so this suggestion must be treated with caution.  Finally, we
note that the LF of NGC 3184 may hint at a cut-off at $\sim
1.7 \times 10^{38} \rm \, ergs \, s^{-1}$, with a couple of
higher luminosity sources skewing the high end of the fit,
but the statistics are limited.
A detailed study of the LF of the
Antennae  has been presented in \citet{zezas01a}.  The
parameters of the LF they derive are very similar to those
presented here.  Also, a study of the LF of M82 has been
presented in \citet{zezas01}. Their slope, although somewhat
flatter is consistent with our results.  

The most striking feature of the results in
Fig.~\ref{fig:lfs} and Table \ref{tab:prop} is that the three
starburst galaxies (NGC 253, M82 and NGC 4038/9) have flatter
luminosity distributions than do the spirals.  This implies
that the starbursts have a larger fraction of higher
luminosity sources relative to the total than do the
spirals.  This can most naturally be explained if systems
with ongoing star formation have a population of
X-ray binaries that dominate the high end of the luminosity
function.  Due to the short time scales of the starbursts,
10-100~Myr, the high luminosity sources are most likely
high-mass X-ray binaries, as the latency period for HMXBs
becoming X-ray luminous is much shorter than for LMXBs.
The hypothesis that the LF slope is related to star formation
rate is supported by the plot in Figure \ref{fig:LF-IR},
which shows the slopes of the luminosity functions of the
galaxies discussed here plotted against the integrated
60$\micron$ luminosity from the IRAS faint source catalog
\citep{moshir92}.  There is a clear correlation (above the 99\%
confidence level) between the
60$\micron$ luminosity -- a measure of the star formation
rate \citep{ken98} -- and the slope of the X-ray luminosity
function, in the sense that flatter slopes have higher star
formation rates.  The outlier is M83, which has a steep slope (deficit
of high luminosity sources) relative to its 60 $\micron$
luminosity. It is worth noting that there are several bright sources
within $D_{25}$ but off the S3 chip which may well be associated with
M83. The slope is therefore an upper limit.

\section{Comparison with Results from Literature}

These results can be compared to studies of early-type
systems in the literature, which are summarized in table~ \ref{tab:lit}.
The elliptical galaxy NGC~4697 has
been studied by \citet{sarazin00}, the S0 galaxy NGC~1553  by
\citet{blanton01}, and the bulge of M31 by  \citet{shirey01}.
The luminosity distributions for NGC~4697, NGC~1553, and the
bulge of M31 are not adequately described by single
power-laws, but instead require broken power-laws.  The
exponents for the high luminosity parts of the cumulative
distributions are $\gamma = 1.79 \pm 0.26$ for M31
\citep{shirey01,primini93}, $\gamma = 1.76^{+1.81}_{-0.39}$
for NGC~4697 \citep{sarazin00}, and $\gamma =
1.7^{+0.7}_{-0.4}$ for NGC~1553 \citep{blanton01}.   For all
three galaxies, the exponent is consistent with the range
$\gamma = 1.5 - 2.1$, steeper than the spirals and starbursts
discussed above.   This suggests that the trend of steeper
slopes correlating with less star formation extends to
early-type spirals and ellipticals.  Two of these galaxies
(NGC 4697 and NGC 1553) are of comparable distance to the
Antennae  and therefore confusion is  potentially a problem. 
The effect of confusion on the luminosity function is
complicated and depends on the clustering and luminosity
properties of the sources.  While a complete analysis of the
effects of confusion is beyond the scope of this paper, we
note that a similarly steep slope is observed in the bulge of
M31 \citep{shirey01,primini93}, lending support to the
hypothesis that early type systems have steeper luminosity
functions.

\section{Model Luminosity Distributions}
\label{sec:models}

As a first step in understanding the luminosity distributions, we adopt
a simple model in which all of the X-ray sources are members of a
single population with uniform properties except for luminosity and
lifetime.  Our goal is to show that this simple model provides an
adequate framework for understanding the difference in luminosity
function slopes between the various types of galaxies and the breaks in
the luminosity functions of individual galaxies.  More detailed
modeling involving the distinctions between high-mass and low-mass
X-ray binaries, and potentially including a distinct source class for
the ultraluminous sources, should be done in the future.  However, the
basic results concerning the luminosity function derived here will
remain true for each individual population of X-ray binaries in a more
complex model.

The number of X-ray point sources is described at each instant in time,
$t$, by the number of sources, $n(t, L, T)$, of luminosity, $L$, and of
age, $T$.  The source birth rate is specified by a function $b(L, t)$
which is, in general, time dependent. The death rate should be
proportional to the number of sources with the constant of
proportionality given by a function $f(L, T)$ which depends on the
source properties, but not explicitly on time.  In general, the
evolution of sources after birth may include luminosity evolution as
well as a luminosity dependent lifetime.  For simplicity, we assume
that each source has a constant luminosity through its lifetime. The
time evolution of $n$ is then

\begin{equation} \frac{d}{dt} n(t,L,T) = b(L,t) \delta(T) - f(L,T)
n(t,L,T) \label{eq:evol} \end{equation}

\noindent where the Dirac delta function $\delta(T)$ enforces the
condition that sources are born with zero age.

The function $f(L, T)$ determines the shape of the source lifetime
distribution.  We choose $f = \delta(T - \tau(L))$.  In this case, each
source of luminosity $L$ lives a time $\tau(L)$ and then promptly
dies.  To specify the lifetime $\tau(L)$, we note that the lifetime of
an X-ray binary is determined by the time required to accrete the
companion star onto the compact object.  Since the luminosity is
proportional to the accretion rate, $L = \eta \dot{M} c^2$ where $\eta
\sim 0.1-0.4$ is the efficiency for conversion of accreted matter to
luminosity, the lifetime will depend on the luminosity as $\tau = \eta
\bar{M}_2 c^2 /\epsilon L$, where $\bar{M}_2$ is the average mass of
the companion stars and $\epsilon$ is the duty cycle of emission
\citep{king01}.  We assume that $\bar{M}_2$ and $\epsilon$ are the same
for all members of the population.

We note that \citet{wu01} presented a model in which the source
population is described only by the number of sources as a function of
$L$ and $t$, $n(t, L)$, i.e.\ no information is retained concerning the
age of the sources.  This is equivalent to choosing a constant for the
function $f$ in our model.  This choice leads to a source lifetime
distribution which is a decaying exponential.  Such a lifetime
distribution has a great spread in ages at fixed luminosity, with many
sources dying immediately after birth and some having extremely long
lifetimes.  Our more general model allows a narrower lifetime
distribution.

We assume a power-law form for the birth rate distribution, $b(L)
\propto L^{-\alpha}$.  The power-law form is purely empirical, and
provides an adequate fit to the data as discussed below.  Further, we
assume that the binaries turn on in X-rays instantaneously after they
are formed.  

If the star formation is dominated by a recent impulsive event, then
the cumulative luminosity distribution shortly after the burst will be
determined by the birth rate distribution,

\begin{equation} N(>L) \propto L^{1-\alpha}. \label{eq:burst}
\end{equation}

If the population is allowed to evolve with no subsequent X-ray binary
formation, then the high end of the luminosity distribution will be
truncated as the highest luminosity sources, with the shortest
lifetimes, die.  The sharpness of the cutoff is determined by the
sharpness of the age distribution.  With the form chosen above for $f$,
the luminosity cutoff will be extremely sharp.  More realistic age
distributions would produce more gradual cutoffs.  Due to the limited
statistics available in constraining such cutoffs, we expect that our
choice for the age distribution is adequate for present data.  In this
case, the differential luminosity distribution is $ n(L) \propto
L^{-\alpha}$ for $ L < L_B$ and $ n(L) = 0 $ for $L \ge L_B$, where the
break luminosity, $L_B$, is determined by the time since the burst,
$t_B$, as $L_B = \eta \bar{M}_2 c^2 /\epsilon t_B$.  The cumulative
luminosity distribution is

\begin{equation} \begin{array}{ll}
N(> L) \propto \left( \frac{L}{L_B}\right)^{1-\alpha} - 1 &
  {\rm ~~ for ~~ } L < L_B  \\
N(>L) = 0  & {\rm ~~ for ~~} L \ge L_B 
\end{array}
\label{eq:agingburst} \end{equation}

Now consider the case of gradual and long-term star formation, leading
to  an equilibrium between X-ray binary formation and death.  If the
lifetimes of the longest lived members of the X-ray point source
population are significantly shorter than interval over which star
formation has been proceeding steadily, then equilibrium will have been
reached.  The death rate at a given $L$ will be proportional to the
number of sources divided by the lifetime $\tau(L)$ and must equal the
birth rate. Since $\tau$ is inversely proportional to $L$
\citep{king01}, the equilibrium luminosity distribution must be
proportional to the birth rate distribution divided by $L$.  The
cumulative luminosity distribution is then

\begin{equation} N(>L) \propto L^{-\alpha}.
\label{eq:continnobreak}
\end{equation}

\noindent This luminosity distribution is steeper than that of the
impulsive case with an exponent that differs by one due to the
(inverse) linear dependence of lifetime on luminosity.

\begin{figure}   \epsscale{0.5} \plotone{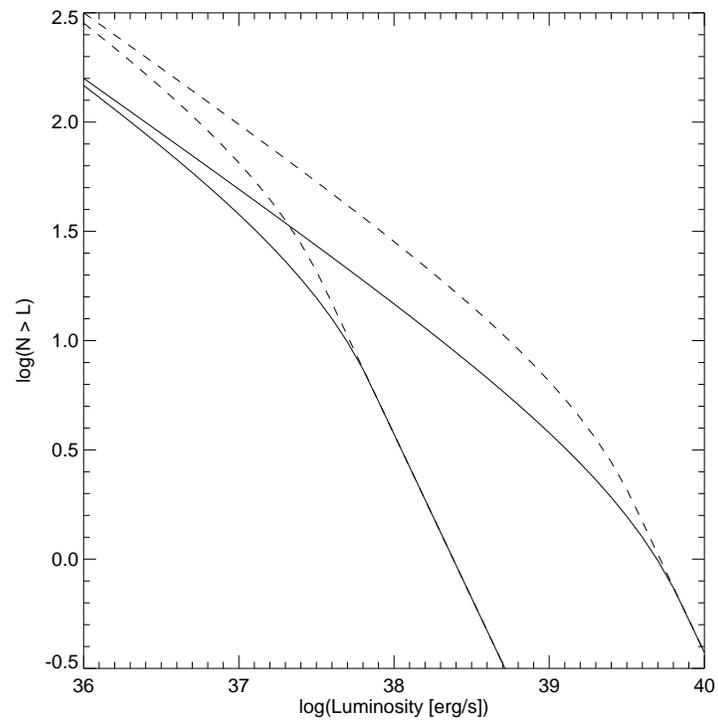}
\caption{Model cumulative luminosity functions.  The upper
two curves are continuous star formation for 10~Myr (solid)
and 20~Myr (dashed).  The lower curves are for 1~Gyr (solid)
and 2~Gyr (dashed) with a star formation rate lower by a
factor of 100. \label{fig:model}}   \end{figure}

At sufficiently low luminosities, the source lifetimes will be longer
than the  interval over which star formation has been proceeding
steadily $t_S$.  In general, the luminosity function will have the form
of Eq.~\ref{eq:burst} for source luminosities corresponding to source
lifetimes longer than $t_B$ and the form of Eq.~\ref{eq:continnobreak}
for higher luminosities.  This leads to a broken power-law form for the
luminosity distribution with the break luminosity, $L_S$, determined by
the time since the burst, $t_S$, with the same relation as between
$L_B$ and $t_B$.  The differential distribution below the break will
have the same slope as that of the birth distribution, while above the
slope will be steeper by one. The cumulative luminosity distribution is
then

\begin{equation} \begin{array}{ll}
N(> L) = \frac{N_0 L_S}{\alpha - 1} 
   \left[ \left( \frac{L}{L_S}\right)^{1-\alpha} - 1 \right] 
   + \frac{N_0 L_S}{\alpha} &
  {\rm ~~ for ~~ } L < L_S  \\
N(> L)  = \frac{N_0 L_S}{\alpha} \left( \frac{L}{L_S}\right)^{-\alpha} 
  & {\rm ~~ for ~~} L \ge L_S 
\end{array}
\label{eq:contin} \end{equation}

\noindent where $N_0$ is the normalization of the different luminosity
distribution at $L_S$.  Plots of this distribution for several values
of $t_S$ are shown in Fig.~\ref{fig:model}.  The older systems have a
steep slope in the high luminosity range.  The younger systems,
analogous to the starbursts, have a flatter slope over the same
luminosity range and distributions extending to higher luminosities. 
The break luminosity depends on the duration over which X-ray binary
formation has been occurring.

If X-ray binary formation proceeds continuously for some interval and
then shuts off, the resulting distribution will be sharply truncated at
high luminosities with the cutoff luminosity determined by the time
since the shut off of X-ray binary formation.  If the duration of the
continuous formation interval is longer than the time since the shut
off of formation, then the (differential) distribution will consist of
two power-laws and a sharp high luminosity cutoff.  If the opposite
situation holds, the luminosity distribution will be as in
Eq.~\ref{eq:agingburst}.

As noted above, we have assumed that all of the X-ray sources are
members of a single population with uniform properties except for
luminosity and lifetime.  In reality two types of X-ray binaries are
known, low-mass and high-mass systems, which have different
evolutionary paths.  High-mass X-ray binaries (HMXBs) are the progeny
of rapidly evolving high mass stars, while low-mass X-ray binaries
(LMXBs) are descendants of slowly evolving low mass stars leading to a
long latency period between star formation and the turn on in X-rays of
the binaries.  In the impulsive star formation case described above,
the long latency of the LMXBs precludes any significant contribution
from LMXBs to the luminosity distribution.  Hence, the X-ray sources in
starbursts are likely HXMBs, as noted above.  There is also likely to
be some delay between the onset of star formation and the appearance of
the first HMXBs, due to time required for the first stars to go
supernova.  However, this can be as short as few million years, so the
assumption made above of instantaneous turn on of the X-ray binaries is
reasonable for star formation episodes with durations in excess of
10~Myr.  

Conversely, the luminosity distribution for old systems is likely to be
dominated by LMXBs because of the much longer lifetimes of the low-mass
companions.  For the equilibrium distributions, LMXBs and HMXBs will
likely have different formation rates, companion masses, and duty
cycles.  This would lead to different luminosity distributions for each
populations, each of which taken individually would be described by the
equations presented above.  The observed distribution would then be the
sum of the distributions for each population.

\section{Conclusions}

Recent Chandra observations show that the X-ray source luminosity
distributions of the early-type galaxies are
steeper than those of the starburst galaxies with an exponent
that differs by approximately one.  This is similar to the
difference between the slopes for our models of a burst of
star formation and steady-state star formation. We suggest
that the luminosity distribution of the starburst galaxies
directly reflects the birth luminosity distribution, while
the other galaxies have a similar birth luminosity
distribution and an observed luminosity distribution modified
by the effects of an aging X-ray binary population. The
slopes of the spirals are intermediate between the two cases,
which may arise from a combination of the two components, as
suggested above for NGC~1291.

For galaxies where the LF is measured over a broad luminosity range, it
may be possible to detect breaks in the luminosity function and relate
them to the galaxy's star formation history \citep{wu01}. 
\citet{shirey01} reports a break in the LF for M31's bulge near $3
\times 10^{37} \rm \, ergs \, s^{-1}$.  Applying the relation found
above between the break luminosity and age of the X-ray binary
population, we find $t_B$ of the order of 10~Gyr for $\epsilon =
0.02$, $\eta = 0.1$, and $\bar{M}_2 = 2 M_{\odot}$.  The difference in
slope above and below the break is $1.32
\pm 0.26$, close to the difference of 1 predicted in our simple model. 
The breaks in NGC~4697 and NGC~1553 are at higher luminosities,
suggesting shorter $t_B$.  However, the break luminosities are
uncertain due to the distance uncertainties.  There are suggestions of
breaks at even higher luminosities, above $10^{39} \rm \, ergs \,
s^{-1}$  in the starbursts M82 and the Antennae.  In the context of the
model presented above, this would be consistent with the rather short
($\lsim 100\, \rm Myr$) durations of the starburst events in these
galaxies.

An alternate explanation of the luminosity function breaks is that they
are due to the transition between neutron star and black hole X-ray
binaries \citep{sarazin00}. This suggestion is motivated by the
observation that the break luminosity in the elliptical galaxies is
close to the Eddington limit of a 1.4 $M_{\odot}$ neutron star
\citep{sarazin00}.  However, the lower break luminosity found in M31
and the higher, although only marginally detected, break luminosities
in the starbursts would argue against this interpretation. 
Additional X-ray luminosity function measurements are
required to determine whether the break luminosity is the
related to the Eddington luminosity or to the star-formation
history of each galaxy as suggested here.

The striking differences in the luminosity functions and the
progress that can be made in understanding the distributions
with a simple model indicate that X-ray point source
luminosity distributions should prove to be a powerful tool
in understanding the evolutionary history of massive star
populations in external galaxies.  New Chandra observations
of a large sample of galaxies across the full Hubble sequence
coupled with detailed stellar evolution models should provide
new insights into the compact object populations of external
galaxies.

\section{Acknowledgements}

We thank the CXC DS and SDS teams for providing the software
used in analysis and Harvey Tananbaum and Vinay Kashyap for valuable comments
on the text and discussions.  This work was partially supported by NASA
contract NAS 8-39073 (CXC) and grant GO1-2092A. PK acknowledges
partial support from NASA grant NAG5-7405.

%\clearpage
\begin{deluxetable}{llrrcccclcc}
\tabletypesize{\scriptsize}
\rotate
\tablecaption{Properties of observations and galaxies \label{tab:prop}}
\tablewidth{0pt}
\tablehead{
  \colhead{Galaxy} & \colhead{Hubble} & \colhead{ObsID} & \colhead{Date} & 
    \colhead{Exposure} & \colhead{$n_{\rm H}$} & \colhead{$D_{25}$ fraction} &
    \colhead{Distance} & \colhead{LF} & \colhead{60 \micron\ luminosity} & \\
  \colhead{} & \colhead{type} & \colhead{} & \colhead{} & 
  \colhead{(sec)} & \colhead{($10^{20}$ cm)} & \colhead{} & 
  \colhead{(Mpc)} & \colhead{Slope} & \colhead{($10^{40}$ erg/s)} }
\startdata
NGC 1291 & S0/a & 795 & 2000-06-27 & 37637 & 2.24 & 0.78 & 8.6 & $1.07 \pm 0.15$ & 8.77 \\
M83 & Sc & 793 & 2000-04-29 & 48562 & 3.70 &  0.73 & 4.7 & $1.38 \pm 0.28$ & 127 \\
NGC 3184 & Scd & 1520 & 2000-01-08 & 38419 & 1.15 &  1.0 & 8.7 & $1.11 \pm 0.22$ & 29.6 \\
IC 5332  & Sd & 2066 & 2001-05-02 & 51986 & 1.38 &  0.81 & 8.4 & $1.30 \pm 0.31$ & 6.09 \\
M 82     & I0 & 361  & 1999-09-20 & 32710 & 4.03 &  0.81 & 5.2 & $0.50 \pm 0.08$ & 1370 \\
NGC 253  & SAB & 969  & 1999-12-16 & 12207 & 1.43 &  0.45 & 3.0 & $0.81 \pm 0.15$ & 324 \\
Antennae & SA/SB(pec) & 315  & 1999-12-01 & 70813 & 3.95 & 0.85 & 25.5 & $0.53 \pm 0.07$ & 1350 \\
\enddata
\end{deluxetable}

\begin{deluxetable}{lllcr}
\tablecaption{LF slope and 60 \micron\ luminosity of early-type galaxies \label{tab:lit}}
\tablewidth{0pt}
\tablehead{
  \colhead{Galaxy} & \colhead {Hubble} & \colhead{LF} & \colhead {60 \micron\ luminosity} &     
    \colhead{Reference}  \\
  \colhead{} & \colhead{type} & \colhead{Slope} & \colhead{($10^{40}$erg/s)} & \colhead{} }
\startdata
NGC 1553 & S0 & $1.7^{+0.7}_{-0.4}$ & 37.7 & \citet{blanton01} \\
NGC 4697 & E6 & $1.76^{+1.81}_{-0.39}$ & 59.5 & \citet{sarazin00} \\
M31 (bulge) & Sb & $1.79 \pm 0.26$ & 102 & \citep{shirey01,primini93} \\
\enddata
\end{deluxetable}

\end{document}